\documentclass[12pt]{article}
\newcommand{\beq}{\begin{equation}\displaystyle}
\newcommand{\eeq}{\end{equation}}
\newcommand{\bit}{\begin{itemize}}
\newcommand{\eit}{\end{itemize}}
\newcommand{\ben}{\begin{enumerate}}
\newcommand{\een}{\end{enumerate}}
\newcommand{\bc}{\begin{center}}
\newcommand{\ec}{\end{center}}

\begin{document} 
\baselineskip 19pt

\bc \large \bf Gain without inversion in a V-type system with low coherence decay rate
for the upper levels.
\ec  \vskip 2.5cm

\noindent {\bf Keywords: V-type three level system, Electromagnetically Induced Transparency,
Lasing without inversion, Atomic Coherence, Autler Townes effect}  \vskip 1.5cm

\noindent {\bf PACS number(s): 42.50.Gy, 42.50.Hz, 42.50.Fx}  \vskip 3.5cm

\begin{center} \bf\normalsize Dipankar Bhattacharyya\footnote[2]
{Permanent address: Santipur College, P.O.- Santipur, Nadia, W.B., INDIA}, Biswajit Ray, 
and Pradip N. Ghosh\,\footnote[1]{Corresponding author.} \\ \,Department of
Physics, University of Calcutta \\ 92, A. P. C. Road, Calcutta --
700 009, \\ I N D I A \\ (Phone : +91 33 2350 8386, Fax : +91 33
2351 9755, \\ E--mail : png@cubmb.ernet.in) 
\end{center}

\newpage\begin{abstract}\baselineskip 22pt
Inversionless gain is observed in a V-type inhomogeneously broadened system without 
introducing any incoherent pumping and only by changing the collisional dephasing decay rate.
In this system sub-Doppler linewidth is achieved with off-resonance pump detuning.

\end{abstract}

\newpage \noindent {\bf I. Introduction:} \vskip 4.5mm

Over the last few years, atomic coherence effect such as coherent
population trapping(CPT) [1,2], electromagnetically induced transparency(EIT)
[3-5], gain without inversion(GWI) and its companion, lasing without inversion
(LWI) [6-8], enhancement of refractive index [9,10], line narrowing [11] etc
have been studied extensively both theoretically and experimentally
due to their multiple applications in various fields. The area of application
has a range from laser cooling to isotope separation and from
ultrahigh-sensitive magnetometers to slowing down of light pulse [12,13]. One of
the most potential applications of the atomic coherence is to extend the
conventional laser sources to ultra-violet and possibly X-rays and even
Gamma-ray spectral range, where the conventional methods based on population
inversion are not available or are difficult to implement. This is performed through
the conversion of incoherent energy into coherent light by a technique known
as LWI [6]. In any commonly considered basis the amplification without inversion
occurs due to the quantum interference between the two dressed-states created by the
strong pump field [14]. The line shape of the probe absorption of a three level system
is controlled by the four different contributions, among them two are absorptive and other
two are dispersive contributions [15]. The absorptive contribution is always positive
where as the dispersive contribution is positive or negative depending upon the type of the scheme
considered.

     Most of the theoretical and experimental work on EIT and LWI are performed
on three level systems(TLS) in the V, lambda($\Lambda$) and cascade($\Xi$) schemes. This is because
these types of systems are easily available in various atomic samples such as Rb, Na, Cs etc.
In recent years, EIT in Doppler broadened media has been observed experimentally using
cw lasers as coupling lasers [16-18]. Recently Ahufinger et al [19] proposed an experiment for obtaining
0.4 $\%$ inversionless gain in a cascade type TLS system applicable to cold free $^{87}$Rb
atoms. Wu et al [20] have studied gain with or without inversion in a V-type system with 
two near degenerate excited levels driven by a strong pump and a weak probe field. They
found that, due to quantum interference between two spontaneous decay channels, even in the
absence of an incoherent pumping the probe gain is achieved.
Boon et al [21] predicted an inversionless gain in a V-type mismatched Doppler-broadened
medium and made a comparison on inversionless gain between matched and mismatched
systems.  
In this article, we report gain without inversion in a Doppler-broadened 
V-type system without introducing any incoherent pumping. An analytical
expression(first order in probe Rabi-frequency) of probe transition is obtained by solving the
density matrix equations. We observed the gain without inversion via electromagnetically 
induced transparency in a Doppler broadened system. We observe that
the significant EIT window is produced for a pump Rabi-frequency greater than
the homogeneous but much less than the Doppler width. We also observe
that EIT window crosses the zero line (Fig.\,4(a)) with decreasing level
dephasing rate($\Gamma_{32}$). Off resonance pump detuning 
gives linewidth narrowing phenomenon for the V-type system.

\noindent {\bf II. Theory:} \vskip 4.5mm

   We consider a V-type three level system (T.L.S) having one common ground level
$|1>$ and two upper levels $|2>$ and $|3>$ as shown in Fig.\,1 with energies
${\hbar}{\omega_{1}}$, ${\hbar}{\omega_{2}}$ and ${\hbar}{\omega_{3}}$ respectively.
$|1> \rightarrow |2>$ and $|1> \rightarrow |3>$ are two dipole allowed transitions.
The strong coherent pump or control field of frequency $\Omega_{1}$ with
electric field amplitude $\varepsilon_{1}$ couples the $|1> \rightarrow |2>$ transition.
A weak probe field of frequency $\Omega_{2}$ with electric field amplitude
$\varepsilon_{2}$ couples the $|1> \rightarrow |3>$ transition whose
dispersion and absorption signal we are interested in. The control and probe
Rabi-frequencies are defined as $\chi_{1}$ = $\frac{\varepsilon_{1}\mu_{12}}{\hbar}$
and $\chi_{2}$ = $\frac{\varepsilon_{2}\mu_{13}}{\hbar}$, where $\mu_{12}$ and
$\mu_{13}$ are the electric dipole moment of the two allowed transitions.
The upper level($|3>$) decays to level $|2>$(due to collision) and $|1>$ with
decay rates $\Gamma_{32}$ and $\Gamma_{31}$. The level $|2>$ decays to $|1>$
with a decay rate $\Gamma_{21}$. The interaction Hamiltonian of the atom and
two fields can be written as
\begin{eqnarray}
      H_{I} = -\frac{\hbar}{2}[ \chi_{1}e^{-i{\Omega_{1}}t}|1><2|
      + \chi_{2}e^{-i{\Omega_{2}}t}|1><3|]  + H.C. 
\end{eqnarray}

The density-matrix equations of motion may be obtained from Liouville
equations under rotating wave approximation [22] we obtained the following
density-matrix equations;

\begin{eqnarray}
   \rho^{.}_{11} = i[\chi_{1}(\rho_{21} - \rho_{12}) + \chi_{2}(\rho_{31} - \rho_{13})]
    + \Gamma_{21}\rho_{22} + \Gamma_{31}\rho_{33}
\end{eqnarray}
\begin{eqnarray}
   \rho^{.}_{22} = i[\chi_{1}(\rho_{12} - \rho_{21})] - \Gamma_{21}\rho_{22} + \Gamma_{32}\rho_{33}
\end{eqnarray}
\begin{eqnarray}
   \rho^{.}_{33} = i[\chi_{2}(\rho_{13} - \rho_{31})] - \Gamma_{31}\rho_{33} - \Gamma_{32}\rho_{33}
\end{eqnarray}
\begin{eqnarray}
   \rho^{.}_{12} = i[\chi_{1}(\rho_{22} - \rho_{11}) + \chi_{2}\rho_{32}]
   - \rho_{12}(\gamma_{12} + i\Delta_{1})
\end{eqnarray}
\begin{eqnarray}
   \rho^{.}_{13} = i[\chi_{2}(\rho_{33} - \rho_{11}) + \chi_{1}\rho_{23}]
   - \rho_{13}(\gamma_{13} + i\Delta_{2}))
\end{eqnarray}
\begin{eqnarray}
   \rho^{.}_{23} = i[\chi_{1}\rho_{13} - \chi_{2}\rho_{21}]
   - \rho_{23}(\gamma_{23} - i(\Delta_{1} - \Delta_{2}))
\end{eqnarray}
and 
\begin{eqnarray}
   \rho_{11} + \rho_{22} + \rho_{33} = 1
\end{eqnarray}

where $\gamma_{12}$, $\gamma_{13}$ and $\gamma_{23}$ are decay rates for the off-diagonal
elements between $|1>$ and $|2>$, $|1>$ and $|3>$ and $|2>$ and $|3>$ respectively.
Off-diagonal or coherence decay rates($\gamma_{ij}$) can be written in terms of
the level or population decay rates($\Gamma_{ij}$)[23,24] as 
$\gamma_{13} = \frac{(\Gamma_{31} + \Gamma_{32})}{2}$, 
$\gamma_{12} = \frac{(\Gamma_{21} + \Gamma_{32})}{2}$ and $\gamma_{23}
= \frac{\Gamma_{32}}{2}$.
$\Delta_{1} = (\omega_{12} - \Omega_{1})$ and $\Delta_{2} = (\omega_{13} - \Omega_{2})$
are the detuning of the pump and probe field where $\omega_{12}$ and $\omega_{13}$
are the transition frequencies corresponding to $|1> \rightarrow |2>$
and $|1> \rightarrow |3>$ transitions.
The dispersion and absorption signal are determined from
$\rho_{13}$(first order in $\chi_{2}$) and keeping $\chi_{1}$ to all orders. Solving equations (2-8)
in steady state we obtain
\begin{eqnarray}
   {\rho}^{1}_{13} = -i\chi_{2} \frac{[\rho^{o}_{11}(\gamma_{23} - i(\Delta_{1} - \Delta_{2}))
+ i \chi_{1}\rho^{o}_{21}]}{[\chi^{2}_{1} + (\gamma_{13} + i \Delta_{2})
(\gamma_{23} - i(\Delta_{1} - \Delta_{2}))]}
\end{eqnarray}
where $\rho^{o}_{11}$ and $\rho^{o}_{21}$ are the zeroth order contributions,
written as
\begin{eqnarray}
\rho^{o}_{11} = \frac{2 \chi^{2}_{1} \gamma_{12} + \Gamma_{21}(\gamma^{2}_{12}
+ \Delta^{2}_{1})}{\Gamma_{21}(\gamma^{2}_{12} + \Delta^{2}_{1}) + 4\chi^{2}_{1}\gamma_{12}}
\end{eqnarray}
and 
\begin{eqnarray}
\rho^{o}_{21} = \frac{{i} \chi_{1}\Gamma_{21}(\gamma_{12} + {i} \Delta_{1})}{\Gamma_{21}
(\gamma^{2}_{12} + \Delta^{2}_{1}) + 4\chi^{2}_{1}\gamma_{12}}
\end{eqnarray}
Equ(9) gives us the absorption(Im($\rho^{1}_{13}$)) and dispersion(Re($\rho^{1}_{13}$))
signal of the probe transition. At low value of pump Rabi-frequency($\chi_{1}$)
Im($\rho^{1}_{13}$) gives a Lorentzian lineshape corresponding to $|1> \rightarrow |3>$
transition. At higher value of $\chi_{1}$ it will split into two symmetric components
commonly known as A-T doublet [25] when $\Delta_{1}$ = 0.
To study the gain in the system we put
$\Delta_{1}$ = $\Delta_{2}$ = 0 and obtain
\begin{eqnarray}
 Im(\rho^{1}_{13}) =
\frac{\chi^{4}_{1}\Gamma_{21}\gamma_{12} - (2\chi^{4}_{1}\gamma_{21}\gamma_{23}
+ 2\chi^{2}_{1}\gamma_{12}\gamma_{13}\gamma^{2}_{23} + \gamma^{2}_{23}
\gamma_{13}\Gamma_{21}\gamma^{2}_{12})}{(\Gamma_{21}\gamma^{2}_{12} + 4\chi^{2}_{1}
\gamma_{12})(\chi_{1} + \gamma_{13}\gamma_{23})^{2}} 
\end{eqnarray}
So for $\Gamma_{32}$($\gamma_{23}$ = $\frac{\Gamma_{32}}{2}$) $\rightarrow$ 0,
Im($\rho_{13}$) become positive for large value of $\chi_{1}$ leading to gain.

Non zero pump detuning gives
an asymmetry in A-T doublet, the position of the A-T peaks can be found from the
pole structure of the Equ(9). The real part of the pole structure gives the position of the
resonance and the imaginary part represents the resonance width.
The position of the resonances are
\begin{eqnarray}
\Delta_{2} = \frac{\Delta_{1}}{2} \pm
\frac{1}{2}\sqrt{\Delta^{2}_{1} + 4\chi^{2}_{1}}
\end{eqnarray}
and the corresponding linewidths are
\begin{eqnarray}
\Gamma = \frac{\Gamma_{31}+ 2\Gamma_{32}}{4}[1 \mp 
\frac{\Delta_{1}\Gamma_{31}}
{\sqrt{\Delta^{2}_{1} + 4\chi^{2}_{1}}(\Gamma_{31} + 2\Gamma_{32})}]
\end{eqnarray}
So for $\Delta_{1} >> \chi_{1}$ the peak at 
$\Delta_{2} = (\frac{\Delta_{1}}{2} + \frac{1}{2}\sqrt{\Delta^{2}_{1} + 4\chi^{2}_{1}})$
has a linewidth smaller than the natural linewidth. Where as the other peak at
$\Delta_{2} = (\frac{\Delta_{1}}{2} - \frac{1}{2}\sqrt{\Delta^{2}_{1} + 4\chi^{2}_{1}})$
is broadened more than the natural width so that they are equal to the unperturbed linewidth.
This feature is also valid when the Doppler broadening is taken into account, 
in that case one peak has sub-Doppler width where the other peak is greater than the Doppler width.
Similar result was obtained by Vemuri et al [11] in $\Lambda$ and $cascede$ systems.

To obtain the probe absorption in a Doppler broadened system the Doppler
broadening should be taken into account. For this purpose Equ(9) should be integrated
over the whole velocity range, the velocity distribution is conventionally
taken as Maxwellian. To introduce the velocity of atom we change the pump and probe
detuning as $\Delta_{1}(v) = (\Delta_{1} \pm \frac{v}{c}\Omega_{1})$ and
$\Delta_{2}(v) = (\Delta_{2} \pm \frac{v}{c}\Omega_{2})$ where $\pm$ sign
indicate the co and counter propagating pump and probe field propagation. We
did our numerical calculation with co-propagating probe and pump beam.
If the number of atoms with velocity $v$ per unit volume is N($v$)d$v$, then
the absorption of the probe laser is
\begin{eqnarray}
\alpha = \int_{-\infty}^{+\infty} Im({\rho^{1}}_{13}) \times (\frac{1}
{u\sqrt{\pi}})exp(-\frac{v^{2}}{u^{2}})dv
\end{eqnarray}
Dispersion can be found similarly,
\begin{eqnarray}
\beta = \int_{-\infty}^{+\infty} Re({\rho^{1}}_{13}) \times (\frac{1}
{u\sqrt{\pi}})exp(-\frac{v^{2}}{u^{2}})dv
\end{eqnarray}
where u  = $\sqrt{\frac{2RT}{M}}$ denote the most probable
velocity of the atom, and it is related to the Doppler width (W$_{d}$) of the probe
spectrum as W$_{d}$ =  ($\omega_{13}$$\sqrt{ln2})\frac{v}{c}$.

\noindent {\bf III. Numerical calculation and results:} \vskip 4.5mm

We study the absorption and dispersion signals of the
probe transition in a Doppler broadened background
for various values of (i) pump Rabi-frequencies($\chi_{1}$) (ii) Doppler
width(W$_{d}$ HWHM) and (iii) $\Gamma_{32}$. In Fig.\,2 Electromagnetically
Induced Transparency(EIT) is observed in a Doppler broadened
back-ground with increasing pump Rabi-frequency.
For numerical calculation in Fig.\,2
we take the values of all population decay rates($\Gamma_{ij}$) as equal to
6 MHz and W$_{d}$ = 280 MHz and in all the figures the probe Rabi-frequency($\chi_{2}$)
is taken as 0.6 MHz.
Here pump field is held on-resonance($\Delta_{1}$ = 0) with $|1>$ 
$\rightarrow$ $|2>$ transition. At zero value of pump field($\chi_{1}$ = 0) a Doppler
broadened probe absorption signal is obtained(dotted curve in Fig.\,2(a)).
With increasing pump field intensity the absorption of the probe signal on the 
line center vanishes(solid curve of Fig.\,2(a))
due to the quantum interference between the dressed states created by strong pump field[4].
As the probe absorption is cancelled in the line center the probe dispersion is
increasing(solid curve of Fig.\,2(b)) that means the system becomes more dispersive the
phenomenon is known as EIT.

For non zero pump detuning($\Delta_{1}$ $\ne$ 0) peak at
$\Delta_{2} = (\frac{\Delta_{1}}{2} + \frac{1}{2}\sqrt{\Delta^{2}_{1}
+ 4\chi^{2}_{1}})$ is reduced in linewidth by a factor
$[ 1 - \frac{\Delta_{1}\Gamma_{31}}
{\sqrt{\Delta^{2}_{1} + 4\chi^{2}_{1}}(\Gamma_{31} +
2\Gamma_{32})}]$ so it becomes narrow, while the other peak at
$\Delta_{2} = (\frac{\Delta_{1}}{2} - \frac{1}{2}\sqrt{\Delta^{2}_{1}
+ 4\chi^{2}_{1}})$ is broadened by the same factor. We study this phenomena
for various value of Doppler width(W$_{d}$) starting from 20 MHz
(dotted curve of Fig.\,3(a)) to 280 MHz(solid curve of Fig.\,3(a))
with a off-resonance pump detuning($\Delta_{1}$) equal to 100 MHz and $\chi_{1}$ = 60 MHz.

   For a pump field Rabi-frequency($\chi_{1}$) = 60 MHz on-resonance
with $|1>$ to $|2>$ transition ($\Delta_{1}$ = 0) and a Doppler width(W$_{d}$)
of 280 MHz gain without inversion is achieved in the system(Fig.\,4).
To observe it we decrease the value
of $\Gamma_{32}$ from 6 MHz(dotted curve of Fig.\,4(a))
to zero(solid curve of Fig.\,4(a)) at an interval of 2 MHz.
It is seen from the figure that with reduction of $\Gamma_{32}$
EIT window increases and gain is achieved(equ.12). It is clear that collisional dephasing 
changes the nature of the spectrum. A Rabi-like side band structure
is observed in the gain profile equally spaced about the line center due to the
split of ground state level. This type of splitting was previously 
observed in V-type system [26,27,21]. We study the effect of Doppler broadening on
gain profile and observed that with increasing Doppler width(W$_{d}$) gain is decreases.
The GWI predicted in this work may be observed for sodium D$_{1}$ transition
(S$_{\frac{1}{2}}$ $\rightarrow$ P$_{\frac{3}{2}}$), F = 1 $\rightarrow$ F$^{'}$ = 1, 2
or F = 2 $\rightarrow$ F$^{'}$ = 1, 2 [28] where the decay rate $\Gamma_{32}$ will be
negligible.

\noindent {\bf IV. Conclusions:} \vskip 4.5mm
In this work we have presented the role of collisional dephasing and off-resonance pump
detuning in a V-type Doppler broadened system in presence of a strong pump or 
coupling field. Recently Wu et al [24] observed how the coherent hole burning phenomena in a $\Lambda$-type 
Doppler broadened system change with atomic collision rate. Here we observed the
probe transmission get amplified(Im($\rho^{1}_{13}$) $>$ 0) with decreasing $\Gamma_{32}$ 
and the maximum amplification is observed when $\Gamma_{32}$ = 0.
So it is seen that collisional dephasing act to reduce the effects 
of quantum coherence. In a V-type system if the upper levels are closely spaced then we can not ignore
$\Gamma_{32}$ but if they are well spaced then we can ignore it. 

\noindent{\bf Acknowledgement:}
Authors thank the Department of Atomic Energy Government of India for 
award of a research grant.

\noindent {\bf References:}
\ben
\setlength{\itemsep}{0ex plus0.2ex}
\setlength{\parsep}{0.5ex plus0.2ex minus0.1ex}
\baselineskip 24pt

\item E. Arimando, Coherent population trapping in laser spectroscopy,
Prog. Opt.(Amsterdam; Elesvier),{\bf 35}, 257(1996).
\item G. Alzetta, A. Gozzini, L. Moi and G. Orriols, NavoCimento
{\bf 36B}, 5(1976).  
\item J. P. Marangos, J. Mod. Opt. {\bf 45}, 471(1998).
\item S. E. Harris, Phys. Rev. Lett. {\bf 62}, 1033(1989).
\item S. E. Harris, Phys. Today {\bf 50(7)}, 36(1997) and the references there in.
\item A. S. Zibrov, M. D. Lukin, D. E. Nikonov, L. Hollberg, M. O. Scully,
V. L. Velichansky and H. G. Robinson, Phys. Rev. Lett. {\bf 75}, 1499(1995).
\item A. Karawajczyk and J. Zakarzewski, Phys. Rev. A {\bf 52}, 1060(1996).
\item J. Monpart and R. Corbal$\acute{a}$n, J. Opt. B: Quant. Semi. Opt. {\bf 2},
R7(2000).
\item M. O. Scully, Phys. Rev. Lett. {\bf 67}, 1855(1991). 
\item M. O. Scully and S. Y. Zhu, Opt. Commun. {\bf 87}, 134(1992). 
\item G. Vemuri, G. S. Agarwal and B. D. N. Rao, Phys. Rev. A {\bf 53}, 2842(1996).
\item L. V. Hau, S. E. Harris, Z. Dulton and C. H. Behoozi, Nature(London) {\bf 397}, 
594(1999).
\item M. S. Bigelow, N. N. Lepeshkin and R. W. Boyd, Phys. Rev. Lett. {\bf 90}, 113903
(2003).
\item J. Mompart and R. Corbal$\acute{a}$n, Opt. Commun. {\bf 156}, 133(1998).
\item G. S. Agarwal, Phys. Rev. A {\bf 55}, 2467(1997).
\item J. Gea-Banacloche, Y Q Li, S Z Jin and M. Xiao, Phys. Rev. A {\bf 51}, 576(1995).
\item A. M. Akulshin, S. Barreiro and A. Lezema, Phys. Rev. A {\bf 57}, 2996(1998).
\item Y. Zhu and T. N. Wasserlauf {\bf 54}, 3653(1996).
\item V. Ahufinger and R. Corbal$\acute{a}$n, J. Opt. B: Quant. Semi. Opt. {\bf 5},
268(2003).
\item J.-H. Wu, Z.-L. Yu and J.-Y. Gao, Opt. Commun. {\bf 211}, 257(2002).
\item J. R. Boon, E. Zekou, D. McGolin and M. H. Dunn, Phys. Rev. A
{\bf 58}, 2560(1998).
\item "Quantum Optics" by M. O. Scully and M. S. Zubairy, (Cambridge University Press London 1997).
\item M. O. Scully and S. -Y. Zhu, Opt. Commun. {\bf 87}, 134(1992).
\item  J.-H. Wu, X.-G. Wei, D.-F. Wang, Y. Chen and J.-Y. Gao,
J. Opt. B: Quant. Semi. Opt. {\bf 6}, 54(2004).
\item S. H. Autler and C. H. Townes, Phys. Rev. {\bf 100}, 703(1955).
\item L. M. Narducci, M. O. Scully, C. H. Keitel and S. -Y. Zhu, Opt. Commun. {\bf 86},
324(1991).
\item W. Tan, W. Lu and R. G. Harrison, Phys. Rev. A {\bf 46}, R3613(1992).
\item V. Wong, R. W. Boyd, C. R. Stroud, R. S. Bennink and A. M. Marino,
Phys. Rev. A {\bf 68}, 012502(2003).
\een

\newpage 
\thispagestyle{empty}
\noindent{\bf $\:\,$Figure Captions} 
\newcounter{fig}
\begin{list}{\bf Figure \arabic{fig} :}{\usecounter{fig}
\setlength{\labelwidth}{2.cm} 
\setlength{\leftmargin}{1.37cm}
\setlength{\labelsep}{0.25cm} 
\setlength{\rightmargin}{0.1mm}
\setlength{\parsep}{0.5ex plus0.2ex minus0.1ex}
\setlength{\itemsep}{0ex plus0.2ex}} \baselineskip 21pt

\item Schematic diagram of a Doppler broadened three-level V-type atomic system driven by
two coherent fields.
\item Probe absorption(a) and dispersion(b) signal in a Doppler broadened
back ground with probe detuning($\Delta_{2}$) for various pump Rabi-frequency($\chi_{1}$).
Other parameters are W$_{d}$ = 280 MHz  $\Gamma_{ij}$ = 6 MHz, 
$\chi_{2}$ = 0.6 MHz and $\Delta_{1}$ = 0.
\item
Probe absorption(a) and dispersion(b) signal with probe detuning for various value
of W$_{d}$. Other parameters are $\chi_{1}$ = 60 and $\chi_{2}$ = 0.6 MHz, 
$\Gamma_{ij}$ = 6 MHz and $\Delta_{1}$ = 100MHz.
\item
Probe absorption(a) and dispersion(b) signal with probe detuning for various value
of  $\Gamma_{32}$. Other parameters are W$_{d}$ = 280 MHz, $\chi_{1}$ = 60 
and $\chi_{2}$ = 0.6 MHz, $\Gamma_{21}$ = $\Gamma_{31}$ = 6 MHz 
and $\Delta_{1}$ = 0.

\end{list}
\end{document}